\documentclass[a4paper,fleqn]{cas-dc}
\usepackage[authoryear,longnamesfirst]{natbib}

\def\tsc#1{\csdef{#1}{\textsc{\lowercase{#1}}\xspace}}
\tsc{WGM}
\tsc{QE}

\newcommand{\dd}{\mathrm{d}}
\newcommand{\csavg}{\langle c_s^2 \rangle}
\newcommand{\cs}{c_s^2}
\newcommand{\eps}{\epsilon}

\begin{document}
\let\WriteBookmarks\relax
\def\floatpagepagefraction{1}
\def\textpagefraction{.001}

\shorttitle{Conformality Thresholds in Neutron Stars}    
\shortauthors{M.~Marczenko}  
\title [mode = title]{Conformality Thresholds in Neutron Stars}  
\author{Michał Marczenko}[orcid=0000-0003-2815-0564]
\ead{michal.marczenko@uwr.edu.pl}
\affiliation{organization={Incubator of Scientific Excellence - Centre for Simulations of Superdense Fluids, University of Wroclaw},
            addressline={plac Maksa Borna 9}, 
            city={Wroclaw},
            postcode={50-204}, 
            country={Poland}}

\begin{abstract}
The dense-matter equation of state (EOS) under neutron-star conditions is investigated. We focus on the qualitative behavior of the speed of sound and its average. We characterize and compare the theoretical criteria for identifying nearly conformal matter. We find that the consistency of these criteria in determining the onset of conformal matter is dictated by the combined state-of-the-art theoretical and astrophysical constraints.
\end{abstract}

\maketitle

\section{Introduction}

Central densities of compact stellar objects exceed the nuclear saturation density ($n_{\rm sat} \simeq0.16~\rm fm^{-3}$). The heaviest NSs have masses exceeding $2~M_\odot$~\citep{Romani:2022jhd}, with densities that reach $6-9~n_{\rm sat}$ at their centers~\citep{Marczenko:2023txe}. Such objects serve as excellent and unique extraterrestrial laboratories of extremely dense matter. The modern observatories for measuring the masses, radii, and tidal deformabilities of compact objects, the gravitational wave interferometers of the LIGO/Virgo Collaboration (LVC)~\citep{LIGOScientific:2018cki, LIGOScientific:2018hze} and the X-ray observatory Neutron star Interior Composition Explorer (NICER) provide new strong constraints on the mass-radius profile of neutron stars~\citep{Riley:2019yda, Miller:2019cac, Miller:2021qha, Riley:2021pdl}. These stringent constraints unleash the potential of NSs as probes of extremely dense matter, allowing for a more systematic study of the influence of the formation of different degrees of freedom within the cores of NSs. Modeling compact stellar objects requires knowledge of the equation of state (EOS) of quantum chromodynamics (QCD). Major progress in constraining the EOS has recently been made by systematic analyses of recent astrophysical observations of the massive pulsar PSRJ0740+6620~\citep{Cromartie:2019kug, Fonseca:2021wxt, Miller:2021qha, Riley:2021pdl} and PSR J0030+0451~\citep{Miller:2019cac}, and the constraint from the recent GW170817 event~\citep{LIGOScientific:2018cki}, within parametric models of the EoS~(see, e.g.,~\citep{Alford:2013aca, Alford:2017qgh, Annala:2017llu, Annala:2019puf}).

The speed of sound provides valuable information about the microscopic description of matter. At low temperature, the low-density EOS ($n_B \lesssim 2~n_{\rm sat}$) is reliably provided by chiral effective field theory~\citep{Tews:2018kmu, Hebeler:2013nza, Drischler:2017wtt, Drischler:2020fvz,Drischler:2020hwi,Keller:2022crb,Drischler:2020yad} where $c_s^2$ is found to be below the conformal value of $1/3$. Asymptotically, $c_s^2$ approaches the conformal limit as dictated by the perturbative QCD (pQCD) calculations~\citep{Gorda:2023mkk, Gorda:2021znl, Gorda:2021gha, Kurkela:2016was, Fraga:2013qra, Kurkela:2009gj, Gorda:2018gpy}. However, the density range found in the NSs is not accessible by any of these methods. Knowledge of the EOS between these two extreme limits comes from inferential analyses of astrophysical observables~\citep{Ecker:2022xxj, Altiparmak:2022bke, Takatsy:2023xzf, Annala:2017llu, Annala:2019puf, Brandes:2022nxa, Brandes:2023hma, Marczenko:2022jhl, Marczenko:2023txe} and effective-model analyses~\citep{Liu:2023ocg, Buballa:2003qv, McLerran:2018hbz, Kovacs:2021ger}. Several suggestions show that the speed of sound exhibits a local maximum at densities realized in the interiors of NSs (see, e.g.,~\cite{Fujimoto:2019hxv, Ecker:2022dlg, Tews:2018kmu}). The structure of the speed of sound remains only hypothetical~\citep{Brandes:2022nxa, Brandes:2023hma}.

The restoration of conformal symmetry inside neutron stars was a subject of recent interest~\citep{Fujimoto:2022ohj} proposed the trace ano\-maly as a signature of conformality. \cite{Annala:2023cwx} proposed another quantity, conformal distance, to delineate the onset of nearly-conformal matter. \cite{Marczenko:2023txe} argued that the onset of conformal matter can be identified with the vanishing curvature of the energy per particle. Recently, it was shown that the ratio of the central pressure to the central energy density can be interpreted as the average speed of sound inside a star (or the average stiffness of the EOS up to the central energy density of a neutron star)~\citep{Saes:2021fzr, Saes:2024xmv} and connected to the trace anomaly and its derivative~\citep{Marczenko:2024uit}. In this work, we represent the conditions for the onset of nearly conformal matter discussed in the literature in terms of the speed of sound and its average value. We demonstrate that they are equivalent under the state-of-the-art theoretical and astrophysical constraints on the equation of state. 

This work is organized as follows. In Sec.~\ref{sec:conf_def}, we introduce thermodynamic quantities related to the speed of sound and conformal symmetry. In Sec.~\ref{sec:disc}, we discuss the differences between various measures of conformality. Sec.~\ref{sec:conc} concludes our findings.

\section{Conformality in neutron stars}
\label{sec:conf_def}

At zero temperature, the speed of sound is defined as
\begin{equation}
    \cs = \frac{\dd p} {\dd \eps} \textrm,
\end{equation}
which is bounded by $0 \leq \cs \leq 1$ due to causality and thermodynamic stability. Assuming that the neutron-star equation of state is smooth and that the pressure is zero at vanishing energy density, one may calculate the average speed of sound~\citep{Saes:2021fzr, Saes:2024xmv, Marczenko:2024uit}:
\begin{equation}\label{eq:csavg}
    \csavg = \frac{1}{\eps} \int\limits_0^\eps \dd \eps\;c_s^2 = \frac{p}{\eps} \textrm,
\end{equation}
which is bounded to $0 \leq \csavg \leq 1$ due to causality and thermodynamic stability. The ratio of the pressure to energy density is related to the energy-density-normalized trace anomaly~\citep{Fujimoto:2022ohj}:
\begin{equation}\label{eq:delta}
    \Delta = \frac{1}{3} - \frac{p}{\eps}\textrm.
\end{equation}
At low density, $\Delta \simeq 1/3$, and it approaches zero as density increases. The trace anomaly was proposed as a signature of conformality in neutron stars~\citep{Fujimoto:2022ohj}. The trace anomaly can be expressed in terms of the average speed of sound using Eq.~\eqref{eq:csavg},
\begin{equation}\label{eq:trace_avg}
    \Delta = \frac{1}{3} - \csavg \textrm.
\end{equation}
Thus, the trace anomaly measures the deviation of the average value of the speed of sound from its conformal value of $1/3$. The relation of the trace anomaly to the speed of sound was recently established~\citep{Fujimoto:2022ohj}:
\begin{equation}\label{eq:cs2_delta}
    \cs = \frac{1}{3} - \Delta - \eps\frac{\dd\Delta}{\dd \eps}
\end{equation}
Inserting Eq.~\eqref{eq:trace_avg} into Eq.~\eqref{eq:cs2_delta} one gets
\begin{equation}\label{eq:cs2_deltap}
    \eps \frac{\dd \Delta}{\dd \eps} = \csavg - \cs \textrm,
\end{equation}
which expresses the rate of change of the trace anomaly in terms of $\cs$ and $\csavg$. That is, if $\cs > \csavg$ ($\cs < \csavg$), the trace anomaly $\Delta$ decreases (increases).

\begin{figure}
    \centering
    \includegraphics[width=1\linewidth]{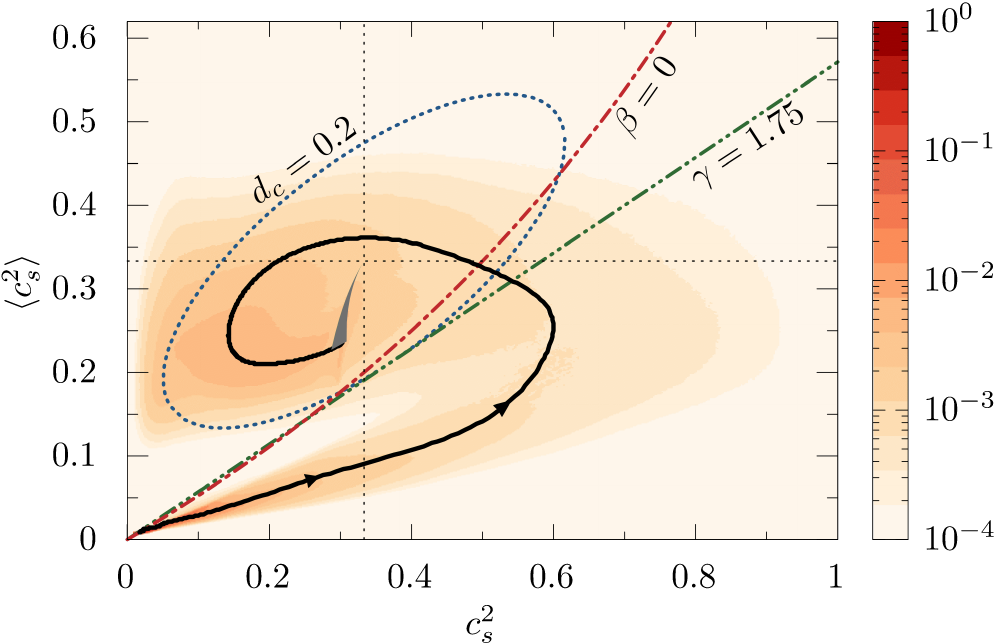}
    \caption{Probability density function (PDF) of the EOS in the $\cs-\csavg$ space. The black, solid line shows the averaged EOS. The arrows on the black line indicate the direction of increasing density. The blue, dotted ellipse shows the $d_c=0.2$ threshold~\citep{Annala:2023cwx}, red, dash-dotted line shows $\beta=0$~\citep{Marczenko:2023txe}, and green, dash-doubly-dotted line marks $\gamma = 1.75$~\citep{Annala:2019puf} threshold. The gray band shows the pQCD constraint. The dashed vertical and horizontal lines mark $\cs = \csavg = 1/3$.}
    \label{fig:measures}
\end{figure}

Another thermodynamic quantity related to the EOS is the polytropic index
\begin{equation}\label{eq:gamma}
    \gamma = \frac{\ln p}{\ln \eps} = \frac{\eps}{p} \cs = \frac{\cs}{\csavg} \textrm,
\end{equation}
where Eqs.~\eqref{eq:trace_avg} and~\eqref{eq:cs2_deltap} were used in the last equality. Typically, at low density, $\gamma \gtrsim 2$, which is typical for a class of relativistic mean-field hadronic EOSs with repulsive vector interactions~\citep{Walecka:1974qa,Serot:1984ey,Serot:1997xg}. \cite{Annala:2019puf} adopted the criterion $\gamma = 1.75$ as a signature for the onset of the conformally symmetric phase.

\cite{Annala:2023cwx} proposed a new quantity to measure the restoration of conformal symmetry,
\begin{equation}
    d_c = \sqrt{\Delta^2 + \left(\eps\frac{\dd \Delta}{\dd \eps}\right)^2} = \sqrt{\left(\frac{1}{3} - \cs\right)^2 + \left(\cs - \csavg\right)^2} \textrm,
\end{equation}
where the last equality holds due to Eqs.~\eqref{eq:trace_avg} and~\eqref{eq:cs2_deltap}. At low density $d_c \simeq 1/3$ and $d_c \rightarrow 0$ as the density increases. The criterion $d_c = 0.2$ was adopted as a delimitation line for nearly conformal matter.

Another interesting observable is the curvature of the energy per particle, $\eps/n$~\citep{Marczenko:2023txe}. The normalized curvature is given by
\begin{equation}
    \beta = \frac{n}{\mu^2} \frac{\dd^2 \eps / n}{\dd n^2} = \cs - 2\frac{1/3 - \Delta}{4/3-\Delta} = \cs - 2\frac{\csavg}{1+\csavg} \textrm,
\end{equation}
where Eq.~\eqref{eq:trace_avg} was used in the last equality. At low density, the energy per particle is a convex function. In the conformal regime, at high density, it is a concave function.  The vanishing curvature $\beta=0$ was proposed to discriminate between the conformally broken and near-conformal regimes. We note that the threshold densities determined by the criteria discussed above ($\gamma=1.75$ and $d_c=0.2$) were found to be consistent with the vanishing curvature of the energy per particle~\citep{Marczenko:2023txe}. 

\section{Equation of State in \texorpdfstring{$\cs-\csavg$}{}~plane}
\label{sec:disc}

The criteria introduced in Sec.~\ref{sec:conf_def} can be expressed in terms of the speed of sound, $\cs$, and its average value, $\csavg$. Therefore, they can be represented in the $\cs-\csavg$ plane. This is depicted in Fig.~\ref{fig:measures}. The average speed of sound is a linear function of the polytropic index (cf.~Eq.~\eqref{eq:gamma}). Thus, the $\gamma=1.75$ criterion is a simple straight line. On the other hand, the curvature of the energy per particle, $\beta=0$, is a more complicated function, and the conformal distance, $d_c=0.2$, forms an ellipse centered around $\cs=\csavg=1/3$. Interestingly, the threshold $\beta=0$ overlaps with $\gamma=1.75$ for small values of $\cs$. This holds because $\beta=0$ when $\gamma = 2/(1+\csavg)$~\citep{Marczenko:2023txe}. Thus, $\gamma \in \langle1;2\rangle$. For $\csavg \simeq 0$, $\gamma$ is close, but less than $2$. This confirms that the value of the polytropic index is close to $\gamma \simeq 2$ at the vanishing curvature of the energy per particle.

\begin{figure}
    \centering
    \includegraphics[width=1\linewidth]{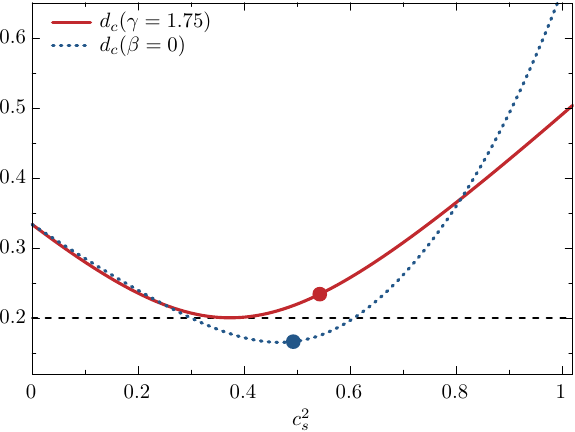}
    \caption{Conformal distance $d_c$ under the constraints $\gamma=1.75$ and $\beta=0$ as a function of the speed of sound. The circles mark the conditions realized by the averaged EOS. The dashed, vertical line shows $d_c=0.2$.}
    \label{fig:measures_as_gamma}
\end{figure}

In Fig.~\ref{fig:measures}, we also show the probability distribution function obtained for a statistical ensemble of EOSs that satisfy the state-of-the-art theoretical and multimessenger constraints (see~\cite{Marczenko:2023txe} for details). The PDF is narrow at small values of $\cs$ and $\csavg$, i.e., low to intermediate densities. This is the effect of the constraints imposed on the EOSs. They cause a rapid increase in the speed of sound at intermediate densities $\eps \sim 3 -5 \eps_{\rm sat}$, while the average speed of sound increases slowly. However, the PDF becomes broader because of the lack of constraints at higher densities. To illustrate the behavior of the PDF better, we consider the averaged EOS. It crosses the thresholds after it develops a maximum $\cs\approx 0.6$. Notably, the thresholds lie in close vicinity of each other.

In Fig.~\ref{fig:measures_as_gamma}, we show the conformal distance, $d_c$, at fixed values of $\gamma=1.75$ and $\beta=0$ as functions of the speed of sound. At vanishing $\cs$, $d_c = 1/3$. At small $\cs$, they overlap (see earlier discussion). They both develop minima around $\cs \approx 0.4 - 0.5$ with $d_c \approx 0.2$. The values of $d_c$ for which the averaged EOS crosses the thresholds are marked by circles. We find that they are close to the demarcation line $d_c=0.2$. This further reaffirms that under the influence of state-of-the-art multi-messenger constraints, the criteria $\gamma=1.75$, $\beta=0$, and $d_c = 0.2$ are equivalent. Therefore, the changeover to the nearly conformal regime is identified with the inflection point of the energy per particle, and thus, its vanishing curvature. 

\begin{figure}
    \centering
    \includegraphics[width=1\linewidth]{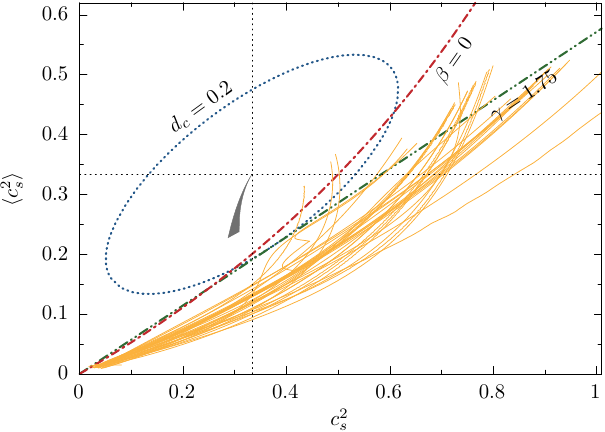}
    \caption{A selection of EOSs (yellow lines) from the~\cite{compose} database in the $\cs-\csavg$ plane shown up to their maximally stable NS configuration. The remainder of the plot is as in Fig.~\ref{fig:measures}.}
    \label{fig:measures_compose}
\end{figure}

Lastly, we show a selection of microscopic hadronic EoSs from the~\cite{compose} database in Fig.~\ref{fig:measures_compose}. We have discarded EoSs that do not fulfill the $M_{\rm max} \geq 2.17~M_\odot$ constraint. The hadronic EoSs are qualitatively in agreement with each other at low densities. We find that most hadronic EoSs do not show conformal behavior. We note that few of the hadronic EoSs that feature $\gamma < 1.75$, $\beta < 0$ also show $d_c < 0.2$, which is consistent with the results shown in~\cite{Annala:2023cwx}.

\section{Conclusion}\label{sec:conc}

In this work, we investigated the cold and dense equation of state under neutron-star conditions. We analyzed observables proposed in the literature to gauge the restoration of conformal symmetry in neutron stars. In particular, we considered the polytropic index~\citep{Annala:2019puf}, curvature of the energy per particle~\citep{Marczenko:2023txe}, and conformal distance~\citep{Annala:2023cwx}. We demonstrated that they admit representation in terms of the speed of sound $\cs$ and its average value $\csavg$. This allowed us to represent them in the $\cs-\csavg$ plane.

We find that the state-of-the-art multi-messenger constraints determine the path of the equation of state in the $\cs-\csavg$ space at low to intermediate densities. Consequently, the EOS crosses all studied conformality thresholds close to each other. Our findings confirm that the changeover to the nearly conformal regime is identified with the inflection point of the energy per particle, and thus, its vanishing curvature. Given the state-of-the-art multi-messenger constraints, we conclude that the vanishing of the curvature of the energy per particle is a sufficient condition for the delineation of the nearly conformal regime at high densities.

\section*{Acknowledgments}
This work was supported by the program Excellence Initiative–Research University of the University of Wroc\l{}aw of the Ministry of Education and Science.

\bibliographystyle{cas-model2-names}
\bibliography{main}

\end{document}